 \documentclass[final,5p]{elsarticle}

\usepackage{graphicx}
\usepackage[usenames,dvipsnames]{color}
\usepackage{adjustbox}
\usepackage{lscape}
\usepackage{booktabs}
\usepackage{rotating}

\usepackage[hyphenbreaks]{breakurl}
\usepackage[hyphens]{url}

\usepackage{amssymb}

\usepackage{hyperref}

\setcitestyle{square}

\journal{Acta Astronautica}

\begin{document}

\begin{frontmatter}



\title{SETI and Democracy}


\author[oxastro]{Peter Hatfield\corref{cor}}
\cortext[cor]{Corresponding author}
\ead{peter.hatfield@physics.ox.ac.uk}


\author[oxlaw,oxworcester]{Leah Trueblood}

\address[oxastro]{Astrophysics, University of Oxford, Denys Wilkinson Building, Keble Road, Oxford, OX1 3RH, UK}
\address[oxlaw]{Faculty of Law, University of Oxford, St Cross Building, St Cross Road, Oxford, OX1 3UL, UK}
\address[oxworcester]{Worcester College, 1 Walton St, Oxford OX1 2HB, UK}

\begin{abstract}

There is a wide-ranging debate about the merits and demerits of searching for, and sending messages to, extraterrestrial intelligences (SETI and METI). There is however reasonable (but not universal) consensus that replying to a message from an extraterrestrial intelligence should not be done unilaterally, without consultation with wider society and the rest of the world. But how should this consultation actually work? In this paper we discuss various ways that decision making in such a scenario could be done democratically, and gain legitimacy. In particular we consider a scientist-led response, a politician-led response, deciding a response using a referendum, and finally using citizens' assemblies. We present the results of a survey of a representative survey of 2,000 people in the UK on how they thought a response should best be determined, and finally discuss parallels to how the public is responding to scientific expertise in the COVID-19 Pandemic.

\end{abstract}

\begin{keyword}
SETI
\end{keyword}

\end{frontmatter}

\parindent=0.5 cm

\section{Introduction}

Search for Extraterrestrial Intelligence (SETI) is the process of searching for intelligent life forms in the Universe beyond planet Earth. There are many such searches underway \cite{Tarter2001}, conventionally using large astronomical telescopes. The current scientific consensus\footnote{At least until very recently\cite{Greaves2020}} is that there is no life (intelligent or otherwise) beyond planet Earth in our Solar System \cite{Lunine2009}; whether there is life elsewhere in the Milky Way or indeed the Universe remains an open question\cite{Brin1983,Lineweaver2002}. Alongside SETI, sometimes researchers consider the prospect of actively Messaging Extraterrestrial Intelligence (METI, sometimes called active SETI) - sending messages out into the Universe with the intention of an as-of-yet unknown intelligence finding and understanding the message\cite{Zaitsev2006}.

Modern SETI projects are generally considered to start with Project Ouzma in 1960\cite{Drake1961,Zuckerman1980}. Notable contemporary examples include Breakthrough Listen\cite{Gajjar2019} and Search for Extraterrestrial Radio Emissions from Nearby Developed Intelligent Populations (SERENDIP)\cite{Werthimer1995}. METI conventionally takes either the form of physical artefacts (e.g. the Voyager Golden Records\cite{Sagan1978}) or electromagnetic signals beamed at astrophysical bodies (e.g. radio signals like the Arecibo Message \cite{Goldsmith2002} or perhaps with pulsed laser signals \cite{Keenan1999}). There is an extensive literature considering how many extraterrestrial species there might be in the Universe \cite{Vakoch2015,Kipping2020}, and the practicalities of sending messages, and how they might be decoded, are fields in and of themselves\cite{Busch2011}. SETI and METI remain speculative enterprises, but have grown as disciplines over time, particularly since the discovery that exoplanets are ubiquitous throughout the Galaxy\cite{Heller2019}.

There is considerable debate in the literature about the merits and demerits of METI\cite{Brin2006,Zaitsev2017,Ball1973,Gertz2016a, Brin2014, Dumas2014,Billingham2014, Denning2010}. Those against typically would argue that METI risks `revealing' the location and existence of our planet, and that it might attract the attention of a hostile species, with potentially civilisation ending consequences. The arguments in favour typically counter that Earth's regular electromagnetic signals (TV etc.) have already revealed our location if that was a reasonable worry, and furthermore that First Contact might bring unimaginable \textit{benefits} if they are friendly\footnote{There are many more details to these arguments, better summarised elsewhere.}. Perhaps the consequences and nature of First Contact would simply be so beyond our current frame of reference that inevitably any predictions will have been made in vain\cite{Denning2011, Denning2013}. The purpose of this paper is not to discuss these arguments, but instead to focus on the issue of \textit{how that decision ought to be made}. Opponents have proposed banning or regulating METI\cite{Billingham2014,Gertz2016b}, so the question is not completely hypothetical - and would rapidly become imperative if there ever was a real life First Contact scenario.  There are a few suggested protocols, but little discussion of how these decision making processes might acquire legitimacy - how can METI be made democratic (if indeed it needs to be).

\section{Previous Democratic Processes in SETI} \label{sec:previous}

Outer space has presented democratic, political, legal and governance issues for decades e.g. regulating satellite ownership and liability and managing ownership issues of mining asteroids. For example, legally a key plank in the international framework for space is the Outer Space Treaty of 1967\cite{OuterSpaceTreaty}, which limits the types of weapon that can be placed in space, forbids governments from making claims over the Moon or other bodies in the Solar System, and regulates ownership of objects launched into space. Politically, exploration of outer space has been charged since the dawn of spaceflight e.g. who is chosen to be an astronaut\cite{Burgess2016,Sage2009,Casper1995}. Here we discuss various democratic processes that have been relevant for SETI and METI to date.

\subsection{Composing Messages} \label{sec:messages}

Early 1970's METI effects were typically led by an individual, or small group of individuals, who then took a series of consultations to improve and clarify their proposed message, before sending. The Pioneer Plaques (launched 1972 and 1973) were essentially developed just by Carl Sagan, Frank Drake and Linda Salzman Sagan\cite{Sagan1972} over three weeks. Similarly the 1974 Arecibo message similarly was developed by a team of about four people led by Frank Drake\cite{CornellSETI}. A few years later for Voyager Golden Records (1977) NASA required a more formal process over the much longer period of a year. The images and music (and message from US President Carter) that went on the discs were selected by a six person committee chaired by Carl Sagan, selected by NASA. NASA is known to have vetoed some choices of the committee, so there was effectively an oversight process\cite{Sagan1978}. Making a plaque/record/message that is representative of all humanity is an incredibly challenging task\cite{Vakoch1998,Gertz2016b}, and the contents of the record have been debated substantially in the subsequent decades\cite{Wolverton2004}.  In all these cases, the number of humans directly contributing to the development of the message was less than ten, with arguably with some informal oversight from a larger number of people from NASA.

IRM Cosmic Call 2 was sent to 5 Sun-like stars in 2003 and contained messages composed by citizens of the USA, Canada and Russia\footnote{\url{https://www.plover.com/misc/Dumas-Dutil/messages.pdf}} \cite{Shuch2011}, likely the first international METI effort. More recent METI attempts have expanded the number of people involved in determining the content of any message to extraterrestrials. In 2008 `A Message From Earth' was sent towards Gliese 581 c, and the content of the message was determined by online submissions and votes on social media website Bebo\footnote{\url{http://news.bbc.co.uk/1/hi/sci/tech/7660449.stm}}. Bebo at the time had 12 million users who in principle could have voted, and half a million actually participated. The trend is clear; by the 2000's METI attempts generally felt that they should try to have some international consultation in the determination of any message\footnote{Also increasingly felt in the literature\cite{Gertz2016a,Billingham2014}}. This trend has continued into the 2010's\footnote{See \cite{Quast2018} for a comprehensive summary of METI messages}, a 2012 follow-up to the `Wow!' signal consisted of 10,000 Twitter messages\footnote{\url{https://www.space.com/17151-alien-wow-signal-response.html}}, and in 2016 the European Space Agency message `A Simple Response to an Elemental Message' included 3775 messages from the worldwide public\footnote{\url{https://blogs.esa.int/artscience/2017/02/17/paul-quast-man-with-a-simple-message/}}.

\subsection{Composing Replies} \label{sec:messages}

In terms of replying to received messages, there is no universally endorsed approach, but probably the most popular/well recognised outlook is the 1989 International Academy of Astronautics SETI Post-Detection Protocols. The protocols lay out some guiding principles to be followed in the event of a detection, and gives an outline of a procedure to be followed, namely to not publicly announce the detection until it has been independently verified, and to then inform other observers through the International Astronomical Union, and the  Secretary General of the United Nations (in accordance with Article XI of the Outer Space Treaty).   In particular the 8th protocol proposes `No response to a signal or other evidence of extra-terrestrial intelligence should be sent until appropriate international consultations have taken place. The procedures for such consultations will be the subject of a separate agreement, declaration or arrangement.'\footnote{\url{https://iaaseti.org/en/declaration-principles-concerning-activities-following-detection/}}\footnote{There are a few revisions and drafted new versions of the Protocol; one modified version instead states `No communication to extraterrestrial intelligence should be sent by any State until appropriate international consultations have taken place. States should not cooperate with attempts to communicate with extraterrestrial intelligence that do not conform to the principles of this Declaration' \url{https://iaaspace.org/wp-content/uploads/iaa/Scientific\%20Activity/setidraft.pdf}} More recently, on the 13th February 2015, SETI experts met at an annual meeting of the American Association for the Advancement of Science, and released a declaration concluding that: `A worldwide scientific, political and humanitarian discussion must occur before any message is sent.'\footnote{\url{https://setiathome.berkeley.edu/meti_statement_0.html}} The question of what amounts to an `international consultation' or a `worldwide scientific political and humanitarian discussion' is left unanswered in the IAA protocol and the AAAS declaration\footnote{Although there are some initial outlines of such a process elsewhere\cite{Brin2014,Billingham2014,Gertz2016b,Michaud1995}}. It is the aim of this paper to discuss what such a process might involve and how it might become to be perceived as legitimate.

\section{Approaches in SETI and Democracy} \label{sec:approaches}

In this section, we will critically assess the range of democratic approaches taken to SETI and METI. We will consider these approaches - scientist-led, representative-led, referendums, and citizens' assemblies - separately as well as the ways in which they are connected. The nature of SETI and METI are such that none of these decision-making methods on their own can satisfy the requirements of the protocol and declaration of a `worldwide scientific political and humanitarian discussion' (Section \ref{sec:previous}).

The key to assessing the appropriateness of the decision-making procedures for SETI and METI is the balance between expertise and judgement. Technocratic, expertise based arguments are more appropriate for those questions where there are clear answers, but for questions of judgement and trade-offs, traditional political representation may be more appropriate\footnote{ `Political issues, by and large, are...not likely to be as arbitrary as a choice between two foods; nor are they likely to be questions of knowledge to which an expert can supply the correct answer. They are questions about action, about what should be done; consequently, they involve both facts and value judgements both ends and means. And, characteristically, the factual judgements and the value commitments, are inexplicably intertwined in political life.'' Hanna Pitkin, The Concept of Representation (University of California 1967). 212 \cite{Pitkin1967}}.  The Protocol and the declaration both suggest with the mention of `political and humanitarian elements' that SETI and METI are not only scientific questions, but also questions of political judgement. Consequently, a range of decision making processes are required for a process to be perceived as legitimate.

\subsection{Scientist/Expert Led}

A scientist led response is probably the most studied mode of reply\cite{Goldsmith1990}, and to some degree is implied by the 9th protocol of the IAA Protocols, which proposes that `...an international committee of scientists and other experts should be established to serve as a focal point for continuing analysis of all observational evidence...' (although it doesn't explicitly suggest that this body should have decision making responsibilities per se). With respect to the Pioneer Plaques and Golden Records, the process was almost entirely expert led. There was some government oversight, via NASA, but the approach to these processes was that this was a fundamentally a scientific question, and so the appropriate method of decision-making was technocratic, based on scientific expertise. The conception of scientific expertise informing these approaches did not involve consulting widely among scientists who took different views, were of different backgrounds, or representing different jurisdictions.

It is important to reflect too on what it means for a scientist to be a representative. Different scientists will likely, and rightly, conceive of their roles in a consultation process of this type differently and come to different conclusions on the science itself\footnote{Weingart gives a helpful account of this in the context of the debate about ozone at 156 \cite{Weingart1999}}.   Not all representatives are elected\footnote{Representation need not mean representative government. A king can represent a nation as can an ambassador. Any public official can sometimes represent the state.  Pitkin 2 \cite{Pitkin1967} },  and there are ways to introduce elements of public engagement and accountability without shifting away entirely from technocratic representation. The question is what the aim of this public engagement is for, is it to act as a check on the actions of representatives or to introduce elements of feedback and transparency.

Consider too the ways in which scientific and political representation can bleed into each other\footnote{`The boundary between politics and science has to be constantly redrawn and reiterated' Weingart 160 \cite{Weingart1999} }.    It is important to think not only what kinds of representatives will be involved in a SETI and METI processes, but how those actors will exercise their expertise and judgement, and to what degree the public will perceive that expertise and judgement as legitimate.

\subsection{Representative Led}

An alternative approach would centre on elected and governmental representatives. To date there has not been huge amounts of interest in playing a major part in SETI by elected representatives at the national government level, but this is of course a mainstay of science fiction e.g. ``Take me to your leader". The IAA protocols do centre the United Nations in the process, and the possibility of international treaties about METI have been considered before\cite{Bilder2020}. The challenge of engaging with the United Nations is that the underlying democratic credentials and political expectations vary enormously from jurisdiction to jurisdiction. Indeed, that kind of approach would bring all of the limitations of achieving international consensus to the process of SETI and METI. A different group of elected representatives, perhaps established specifically for this purpose, would have opposing problems in that while it could be more nimble, it could have far less publicity and so would likely be perceived to be less legitimate. It is important to emphasise again that certain decisions about METI, particularly a ``take me to your leader'' scenario, have a very short turnaround time. It is essential that there is a global discussion in advance\cite{Billingham2014,Denning2019,Gertz2016b} in order to avoid SETI and METI descending into realpolitik\cite{Dominik2011}, and to continue to see SETI and METI in scientific and humanitarian terms rather than defence terms.

\subsection{Direct Democracy/Referendums}

`A Message From Earth' and other similar crowd-sourced METI efforts are of course drawing heavily upon Direct Democracy ideas. SETI@Home, where members of the public can contribute some of their home computing power to SETI searches\cite{Anderson2002}, similarly to some degree draws upon the principle that SETI/METI are a universal endeavour that should be carried out collectively. The challenges of working across languages and political systems persist from the challenges with elected representatives, and inequities with respect to access to technology should be taken seriously too. Even in these cases, there is worry about what the democratic credentials of such a process would be without a global demos, and whether a global demos is possible such that cross-national processes can have democratic legitimacy\footnote{Valentini introduces some of these and challenges them in Laura Valentini, `No Global Demos, No Global Democracy? A Systematization and Critique' (2014) 12 Perspectives on Politics 789 \cite{Valentini2014}}.  Furthermore, the voting stage of a referendum is significantly shaped by the agenda-setting stage and a familiar criticism of these types of processes is that they are dominated by elites\footnote{As Walker rightly argues `political actors use referendums to achieve their goals.' M Walker, The Strategic Use of Referendums: Power, Legitimacy, and Democracy (Springer 2003). 1 }.  If referendums, polls, and other forms of direct democracy are employed, it must be clear in advance how these processes are being used and what the implications are in advance, and there must be equity in the agenda-setting processes that takes different languages, political backgrounds, and scientific perspectives seriously.

Governments occasionally reply to petitions on topics to do with SETI e.g. 17,465 signatures on a petition to the White House that got an official response\cite{Gulyas2013}\footnote{\url{https://www.cbsnews.com/news/white-house-no-evidence-of-space-aliens/}} . There has been one real world government mandated referendum to do with the topic of SETI to our knowledge; Initiative 300 in Denver, Colorado, USA. Voters in the area on the 2nd November 2010 voted on whether or not to charge the city with creating a seven person commission to investigate UFOs. Campaigners had to get 3,973 signatures for the referendum to go ahead, but it was rejected 31,108 (17.66\%) to 145,022 (82.34\%).

\subsection{Consultative/Citizens' Assembly}

Citizens' assemblies, or mini-publics, are processes where participants are selected randomly and participation reflects broader demographics\footnote{Similar in principle to juries.}\cite{Rose2009}.  So, for example, if the population is 52\% female so too would the participants be. This creates challenges for understanding what demographic characteristics to isolate. Citizens' assemblies are sometimes used as freestanding democratic processes, and at other points they are built into larger democratic processes such as by producing proposals that then lead to referendums\cite{Pal2012}.  They provide a reasonably efficient mechanism of public consultation, but not the sort of broad-based public legitimacy of wide-spread participation.

\subsection{Individual Led}

The role of individuals and companies in going to space has been significant \cite{Dula1985,Denis2020}; there have even been a few METI efforts that were advertisements for products\cite{Quast2018}.  This allows for nimbleness but not the sort of widespread public perception of legitimacy that is necessary for the sort of broad consensus imagined by the IAA protocol\cite{Gertz2016b}.

\section{Survey Data} \label{sec:poll_data_summary}

Over 2019-2020 we performed a series of formal and informal projects to sample opinions in the UK on how METI efforts might best acquire democratic legitimacy. These public perceptions are crucial for understanding what kinds of processes have broad public support. The most rigorous of these studies was a survey we commissioned from the British polling agency Survation. On our behalf, Survation asked 2000 people 18 years or older resident in the UK two questions:

\textbf{Question 1)}

\textit{Imagine a scenario in which scientists receive an unambiguous message from extraterrestrials (alien life forms) on a distant planet. Of the following options, which would be your preference in terms of how humanity's response to this message should be determined?}

\begin{enumerate}
    \item Team of scientists		\hfill		39\%
    \item By elected representatives		\hfill			15\%
    \item By a planet-wide referendum			\hfill		11\%
    \item By a citizens' assembly of randomly selected adults	\hfill	\hfill11\%
    \item Don't know				\hfill			23\%
\end{enumerate}

\textbf{Question 2)}

\textit{In the event that a planet-wide referendum on whether to reply to a message from extraterrestrials or not was held, would you vote to initiate contact with the alien species, vote to not initiate contact with the alien species or would you not vote in that referendum?}

\begin{enumerate}
    \item Vote to initiate contact with the alien species		\hfill	56\%
    \item Vote to not initiate contact with the alien species	\hfill	13\%
    \item Would not vote				\hfill		10\%
    \item Don't know					\hfill		21\%
\end{enumerate}

Survation weighted the sample to be representative in terms of age, sex, region, household income, education, 2017 General Election Vote, 2016 EU Referendum Vote, and 2019 European Parliament Election Vote, and we have answers to the poll broken down by those responses. The data was taken 2nd - 5th September 2019. These breakdowns are included in Appendix A, but responses largely were uncorrelated with demographics.

The polling questions were designed to be indicative rather than definitive, and of course Q1 artificially separates the approaches; people largely tried to suggest a mixture when given the option in person. Participants also occasionally brought up the issue of whether a world-wide referendum would be administratively possible. To our minds such an undertaking would indeed be challenging - but 614,684,398 people voted in the 2019 Indian general election, so democracy on a large scale is possible.

To our knowledge no comparable polling for the first question exists. Our second question has been asked in various forms, although to our knowledge this is the first time it has been asked posed in the form of a voting intention question. Our Q2 however is consistent with other polls that find that in the absence of more information people's attitudes are largely positive towards communication with extraterrestrials e.g. a 2015 Yougov UK poll in 46\% found in favour of communication, 33\% against communication and 21\% don't know\footnote{https://yougov.co.uk/topics/lifestyle/articles-reports/2015/09/24/you-are-not-alone-most-people-believe-aliens-exist} .

The limits of the poll are clear and important 1) it covers only the UK (and just 18 year olds and older) and 2) it was taken at an exceptional time for democracy in the UK. These limitations did however in some ways make the question a `Rorschach test' for attitudes to democratic processes. The authors' interactions with people suggested quite a low satisfaction with representatives and referendums (although that is for another time and place). It also proved a useful pedagogical tool to get members of the public to think about different democratic processes in a non-partisan way.

We also asked the question in a range of less structured ways, including: British Science Festival (10th September 2019), Stargazing Oxford (25th January 2020), secondary schools (Europa School in Oxfordshire on the 15th March 2019, Dover Grammar School for Girls in Kent on the 1st November 2019) and a British Science Association twitter poll\footnote{\url{https://twitter.com/BritishSciFest/status/1171422013308559362}}. Interestingly the results were fairly robust across slightly different audiences and slightly different ways of asking the question.

Participants did sometimes get to give their own suggestions of how to make the decision. Suggestions included: letting famous individuals make any decision (Sir David Attenborough and HM Queen Elizabeth II were suggested), flipping a coin, letting psychologists decide and letting children decide\footnote{This possibility had in fact, unknown to the authors, already been implemented in the Teen Age message and the `New Arecibo Message', which both used submissions from youths \cite{Quast2018}.}. 

Finally we did during 2019 ask a real sitting Member of the UK House of Commons (who will remain anonymous) this question, who simply said that they did not want the referendum option.

\section{Recommendations} \label{sec:recommendations}

The purpose of the protocol and the declaration are to say that a greater degree of legitimacy is required for processes associated with SETI and METI. It is no longer enough - if it ever was - for a few scientists to go it alone. It is essential for purposes of legitimacy of the processes that the public sees them as legitimate. This creates a tension in that processes that are genuinely consultative take time, and some instances of METI may happen very quickly.
It is essential too to consider what kind of issues SETI and METI are. We have argued so far that they include elements of expertise and judgement, and so this is going to require different types of representatives working together with the public around the world. 

We broadly support the view put forward by Churchill that scientists should be `on tap, but not on top'. The IAA has rightly signalled that there are scientific issues at stake, but not scientific issues that are beyond doubt, and they intersect too with political issues that are questions of judgement. We think that the process should be driven by a team of scientists nominated by different jurisdictions rather than countries (the global north, global south etc) with broad opportunities for consultation through polls and citizens' assemblies. Ideally there would be a role for elected representatives who already specialise in science such as those who lead parliamentary select committees on science.

Essentially: there is a role for scientists, individuals, referendums, and representatives in the process of designing `global humanitarian and political consultation'. It requires attending to the parts of SETI and METI that are technical scientific issues, how to communicate (which requires social science and humanities as well as science) and those of judgement, should such initiatives be undertaken in the first place, what should be said. The best balance between these different types of representation will lead to the best outcome. Designing such a process is challenging, but when grounded on the right principles it is certainly possible.

\section{The COVID-19 Pandemic as a SETI Event Proxy} \label{sec:education}

The bulk of the research and thinking behind this paper was performed in 2019. In 2020 the world was hit by a real-life scientific crisis that affected the entire world, across society, leading to mass death and economic damage. Here we discuss similarities and differences between the pandemic and a First Contact event, and why the world's response to COVID-19 might be the best chance of seeing how governments and scientists might interact in such a scenario\cite{Traphagan2020}, before a First Contact event actually happens.

Although the pandemic remains an evolving situation at the time of writing, we suggest that there are some similarities between the COVID-19 crisis and a First Contact event for the following reasons - both are:
\begin{itemize}
    \item Crises of a fundamentally scientific nature
    \item Also of huge social, moral, economic, political impact
    \item World-wide crises affecting essentially every human on Earth
    \item `External' threats i.e. all of humanity is `on the same side' (in contrast to say, WWII)
    \item `Out of the blue'; both have been seen as possible in advance, but were essentially unanticipated
\end{itemize}

They are of course also many ways in which a pandemic is different from First Contact:
\begin{itemize}
    \item Pandemics are not completely without precedent; the last major global pandemic was ~100 years ago, whereas First Contact would be completely without precedent
    \item Pandemics are wholly negative events, whereas First Contact could have both positive and negative impacts
\end{itemize}

The full social impact of COVID-19 will take many years to unravel. We briefly note a few salient occurrences that may be insightful for how a real First Contact incident could occur (again predominantly from a UK perspective):

\begin{itemize}
    \item Scrutiny of individual involved scientists skyrockets to levels normally only experienced by top level politicians\footnote{\url{https://www.bbc.co.uk/news/uk-politics-52553229}}
    \item Other scientists setting up alternative sources of advice\footnote{\url{https://www.thetimes.co.uk/article/independent-sage-committees-scientists-have-every-right-to-criticise-the-government-d7rdvcb22}} 
    \item The public rapidly learning that scientific advice can be conflicting\footnote{\url{https://www.bbc.co.uk/news/science-environment-52522460}}
\end{itemize}

In particular we would note the response has been largely led by politicians. For example, politicians could have at the start of the crisis completely passed over judgement about when lockdowns would be imposed and raised to an independent body. This did not happen in the UK (and not to our knowledge anywhere else in the world). Similarly it is not completely unimaginable to completely pass decision making about the crisis out to the public; let people vote every three weeks if they were ready to reduce the severity of lockdown\footnote{Challenging technologically but not beyond the realms of possibility if there had been felt a real need for it.}. There were some weak efforts at direct democracy e.g. petitions, but these currently have far fewer signatures than the petitions about Brexit, which clearly there was much more demand for direct input into the process (the top Brexit-related petition has $\sim$ 6 million signatures, wheras the top COVID-19 related petition only has approximately a tenth of this). There were also some small citizen-assembly-like processes, but essentially there was no desire for the public to be directly involved in decisions about lockdown. All of these are observations about how the crisis played out rather than approval/criticism. Could a committee of just scientists have had the authority to order lockdown? Would voting to end or lift lockdown have led to good outcomes? We will likely never know the answers to these questions - hopefully it will be many years before democracies are given comparably challenging decisions.

\section{Conclusions} \label{sec:conclusions}

While historically consultation processes with respect to SETI and METI have been ad hoc and limited, the IAA have signalled with their declaration that a more demanding process is required going forward. We have surveyed different models of representation in light of the different parts of the METI and SETI questions, which are in fact composed of a number of sub-questions about science and politics. We argued that different types of representation are required for different parts of this question, and that many types of representation will need to work together. We suggested some recommendations for how this is possible. This takes time, and it is important to be proactive rather than reactive given that METI can happen quickly, but it is certainly possible to satisfy the standard of a global humanitarian and political and scientific conversation, and it may be that having such conversations is timely for other global issues such as climate change and public health. 

Key points:

\begin{itemize}
    \item Within the limited scope of this project, it appears the the public are happy about scientists having a key role in determining how contact with extraterrestrials is made
    \item The COVID-19 crisis gives an insight into how scientific advice can be politicised very rapidly, and suggests that scientific and representative mechanisms will find it easy to dominate any First Contact scenario
    \item Thought should be given to making sure dealing with a detection has maximum legitimacy. One possible way we believe this could be achieved is by having decision making driven by a team of scientists nominated by different jurisdictions (rather than nation states) with broad opportunities for consultation; ideally there would be elected representatives who already specialise in science. For example those who lead parliamentary select committees (and similar) on science might be particularly well placed to contribute to decision making, having both expertise as well as democratic legitimacy.

\end{itemize}

\section*{Acknowledgements}

Many thanks to the Oxford Physics Outreach Project Fund for funding the survey. Thanks to Damian Lyons Lowe and Survation for their support and input into this project.
Particular thanks to Prof. Steven Rose (Imperial) and Dr Sian Tedaldi (Oxford) who gave valuable input to the ideas in this article, and to the British Science Festival, Europa School and Dover Grammar School for Girls for giving us an opportunity to explore and develop the ideas in this text.

\appendix
\section[]{Full Survey Data} \label{sec:full_survey_data}

\begin{landscape}
\begin{figure}

\includegraphics[scale=0.7]{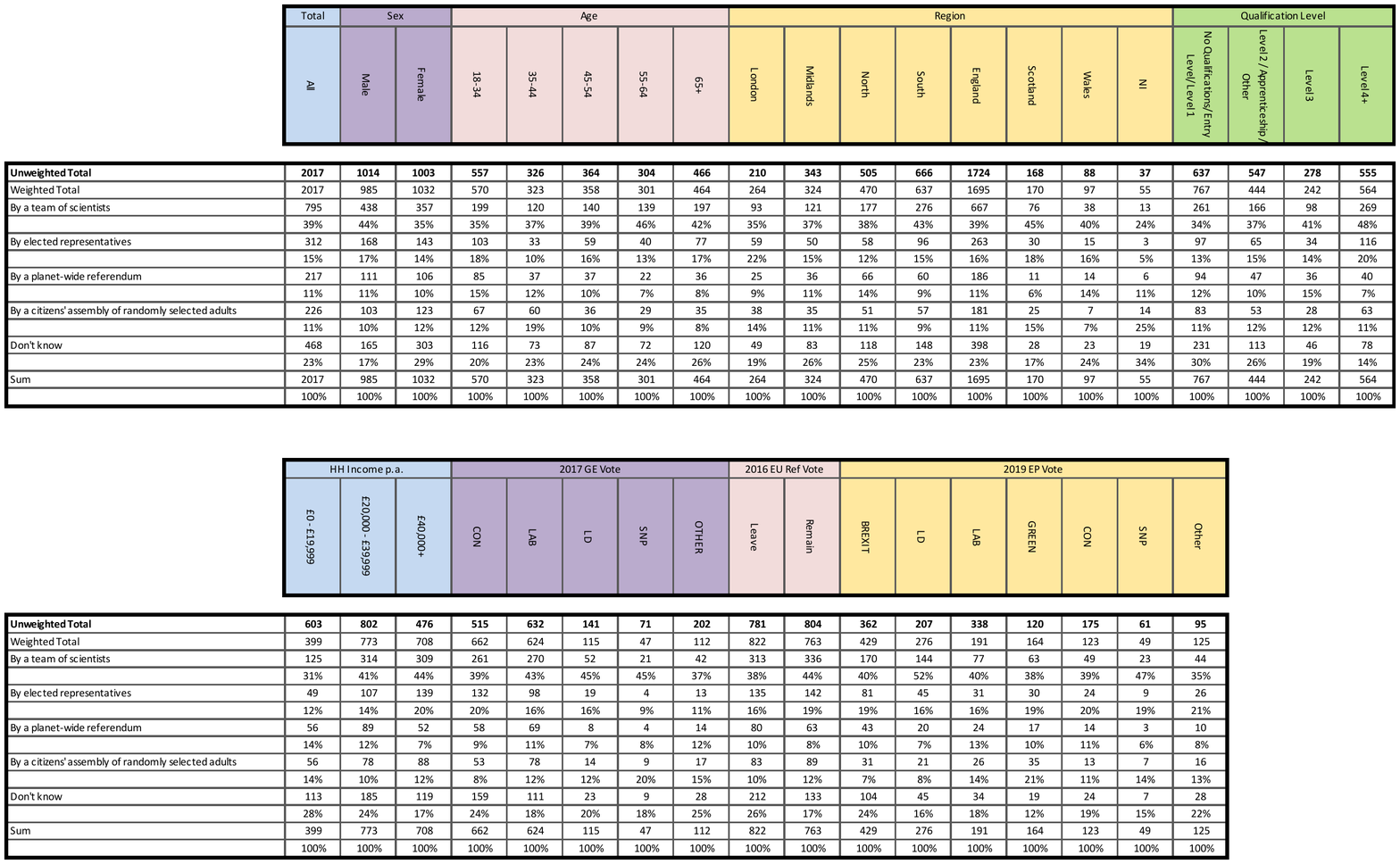}
\caption{The polling results for Question 1: "Imagine a scenario in which scientists receive an unambiguous message from extraterrestrials (alien life forms) on a distant planet. Of the following options, which would be your preference in terms of how humanity's response to this message should be determined?"}
\label{fig:poll1}
\end{figure}
\end{landscape}

\begin{landscape}
\begin{figure}

\includegraphics[scale=0.7]{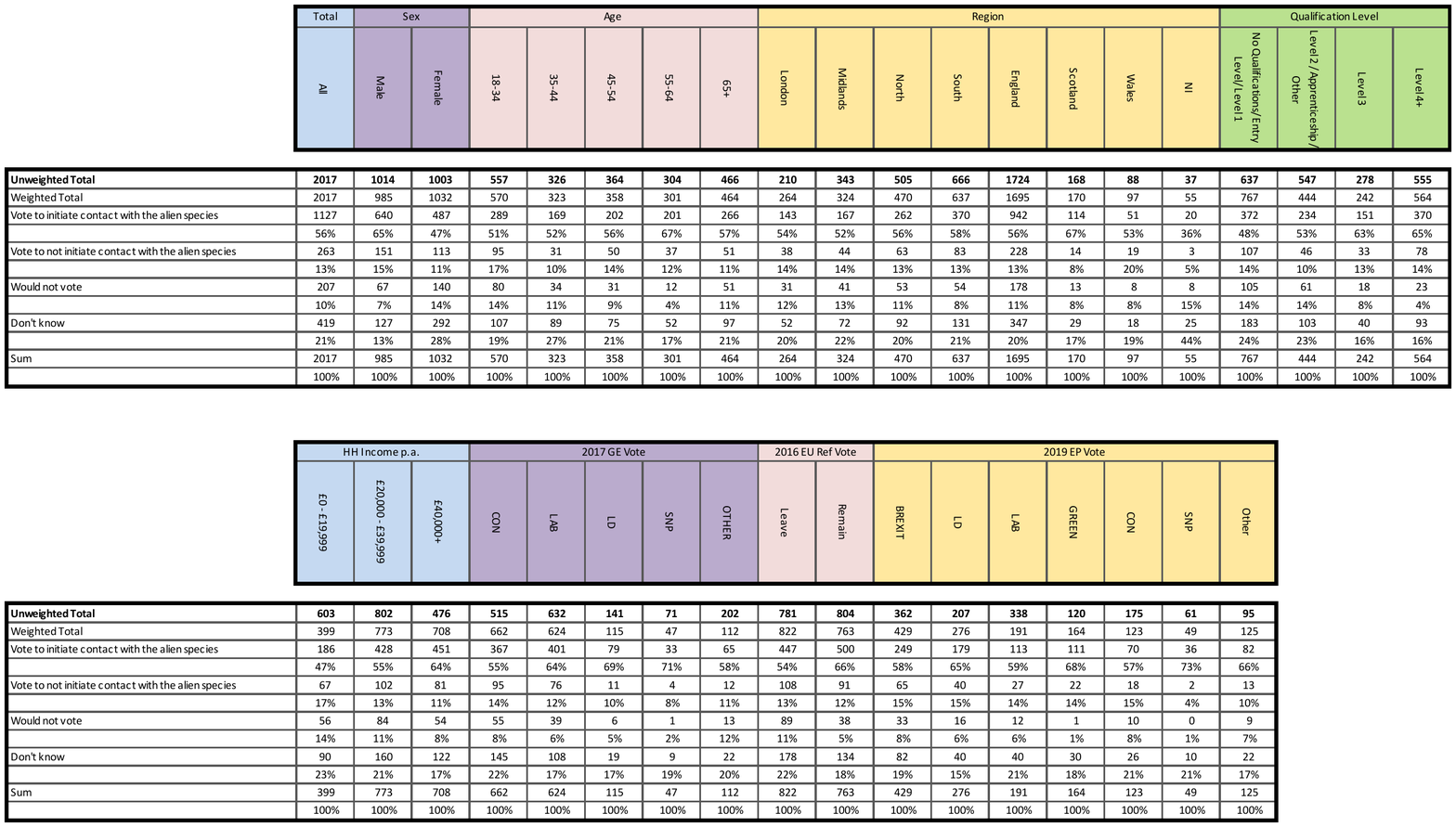}
\caption{The polling results for Question 2: "In the event that a planet-wide referendum on whether to reply to a message from extraterrestrials or not was held, would you vote to initiate contact with the alien species, vote to not initiate contact with the alien species or would you not vote in that referendum?" }
\label{fig:poll2}
\end{figure}
\end{landscape}

Tables \ref{fig:poll1} and \ref{fig:poll2} show the full polling data for Questions 1 and 2 discussed in Section \ref{sec:poll_data_summary}. Data were analysed and weighted by Survation. Survation are members of The British Polling Council and a limited company registered in England and Wales with number 07143509.

The survey was conducted over 2nd-5th September 2019 via online panel. Invitations to complete the survey were sent out to members of the panel, and differential response rates from different demographic groups were taken into account. Data were weighted to the profile of all adults in the UK aged 18 and over. Data were weighted by age, sex, region, household income, education, 2017 General Election Vote (8th June 2017), 2016 EU Referendum Vote (23rd June 2016), and 2019 European Parliament Election Vote (23rd May 2019). Targets for the weighted data were derived from Office for National Statistics data and the results of the respective votes. In all questions where the responses are a list of parties, names or statements, these will typically have been displayed to respondents in a randomised order. The political parties abbreviations are CON (Conservative Party), LAB (Labour Party), LD (Liberal Democrat Party), SNP (Scottish National Party), BREXIT (Brexit Party) and GREEN (Green Party of England and Wales).

The tables show the unweighted total of individuals in a given sub-population, the weighted total, and then the weighted number and percentages of people selecting the given options. Needless to say, not all differences in survey answers between sub-groups are statistically significant.

\clearpage

\textbf{REFERENCES}

\bibliographystyle{elsarticle-num.bst}
\bibliography{seti_democracy}{}

\begin{thebibliography}{10}
\expandafter\ifx\csname url\endcsname\relax
  \def\url#1{\texttt{#1}}\fi
\expandafter\ifx\csname urlprefix\endcsname\relax\def\urlprefix{URL }\fi
\expandafter\ifx\csname href\endcsname\relax
  \def\href#1#2{#2} \def\path#1{#1}\fi

\bibitem{Tarter2001}
J.~Tarter,
  \href{http://www.annualreviews.org/doi/10.1146/annurev.astro.39.1.511}{{The
  Search for Extraterrestrial Intelligence (SETI)}}, Annual Review of Astronomy
  and Astrophysics 39~(1) (2001) 511--548.
\newblock \href {http://dx.doi.org/10.1146/annurev.astro.39.1.511}
  {\path{doi:10.1146/annurev.astro.39.1.511}}.
\newline\urlprefix\url{http://www.annualreviews.org/doi/10.1146/annurev.astro.39.1.511}

\bibitem{Greaves2020}
J.~S. Greaves, A.~M.~S. Richards, W.~Bains, P.~B. Rimmer, H.~Sagawa, D.~L.
  Clements, S.~Seager, J.~J. Petkowski, C.~Sousa-Silva, S.~Ranjan,
  E.~Drabek-Maunder, H.~J. Fraser, A.~Cartwright, I.~Mueller-Wodarg, Z.~Zhan,
  P.~Friberg, I.~Coulson, E.~Lee, J.~Hoge,
  \href{https://ui.adsabs.harvard.edu/abs/2020NatAs.tmp..234G/abstract
  http://www.nature.com/articles/s41550-020-1174-4}{{Phosphine gas in the cloud
  decks of Venus}}, Nature Astronomy\href {http://arxiv.org/abs/2009.06593}
  {\path{arXiv:2009.06593}}, \href
  {http://dx.doi.org/10.1038/s41550-020-1174-4}
  {\path{doi:10.1038/s41550-020-1174-4}}.
\newline\urlprefix\url{https://ui.adsabs.harvard.edu/abs/2020NatAs.tmp..234G/abstract
  http://www.nature.com/articles/s41550-020-1174-4}

\bibitem{Lunine2009}
J.~I. Lunine, \href{https://www.jstor.org/stable/20721510?seq=1}{{Saturn's
  Titan: A Strict Test for Life's Cosmic Ubiquity}}, Proceedings of the
  American Philosophical Society 153~(4) (2009) 403--418.
\newline\urlprefix\url{https://www.jstor.org/stable/20721510?seq=1}

\bibitem{Brin1983}
Brin, G.~D.,
  \href{https://ui.adsabs.harvard.edu/abs/1983QJRAS..24..283B/abstract}{{The
  Quarterly journal of the Royal Astronomical Society.}}, Vol.~24, Published
  for the Royal Astronomical Society by Blackwell Science, 1983.
\newline\urlprefix\url{https://ui.adsabs.harvard.edu/abs/1983QJRAS..24..283B/abstract}

\bibitem{Lineweaver2002}
C.~H. Lineweaver, T.~M. Davis, \href{https://pubmed.ncbi.nlm.nih.gov/12530239/
  http://www.liebertpub.com/doi/10.1089/153110702762027871}{{Does the Rapid
  Appearance of Life on Earth Suggest that Life Is Common in the Universe?}},
  Astrobiology 2~(3) (2002) 293--304.
\newblock \href {http://dx.doi.org/10.1089/153110702762027871}
  {\path{doi:10.1089/153110702762027871}}.
\newline\urlprefix\url{https://pubmed.ncbi.nlm.nih.gov/12530239/
  http://www.liebertpub.com/doi/10.1089/153110702762027871}

\bibitem{Zaitsev2006}
A.~Zaitsev, \href{http://arxiv.org/abs/physics/0610031}{{Messaging to
  Extra-Terrestrial Intelligence}}.
\newline\urlprefix\url{http://arxiv.org/abs/physics/0610031}

\bibitem{Drake1961}
F.~D. Drake,
  \href{http://physicstoday.scitation.org/doi/10.1063/1.3057500}{{Project
  Ozma}}, Physics Today 14~(4) (1961) 40--46.
\newblock \href {http://dx.doi.org/10.1063/1.3057500}
  {\path{doi:10.1063/1.3057500}}.
\newline\urlprefix\url{http://physicstoday.scitation.org/doi/10.1063/1.3057500}

\bibitem{Zuckerman1980}
B.~Zuckerman, J.~Tarter,
  \href{http://link.springer.com/10.1007/978-94-009-9115-6{\_}10}{{Microwave
  Searches in the U.S.A. and Canada}}, Springer, Dordrecht, 1980, pp. 81--92.
\newblock \href {http://dx.doi.org/10.1007/978-94-009-9115-6_10}
  {\path{doi:10.1007/978-94-009-9115-6_10}}.
\newline\urlprefix\url{http://link.springer.com/10.1007/978-94-009-9115-6{\_}10}

\bibitem{Gajjar2019}
V.~Gajjar, A.~Siemion, S.~Croft, B.~Brzycki, M.~Burgay, T.~Carozzi, R.~Concu,
  D.~Czech, D.~DeBoer, J.~DeMarines, J.~Drew, J.~E. Enriquez, J.~Fawcett,
  P.~Gallagher, M.~Garrett, N.~Gizani, G.~Hellbourg, J.~Holder, H.~Isaacson,
  S.~Kudale, B.~Lacki, M.~Lebofsky, D.~Li, D.~H.~E. MacMahon, J.~McCauley,
  A.~Melis, E.~Molinari, P.~Murphy, D.~Perrodin, M.~Pilia, D.~C. Price,
  C.~Webb, D.~Werthimer, D.~Williams, P.~Worden, P.~Zarka, Y.~G. Zhang,
  \href{http://arxiv.org/abs/1907.05519}{{The Breakthrough Listen Search for
  Extraterrestrial Intelligence}}\href {http://arxiv.org/abs/1907.05519}
  {\path{arXiv:1907.05519}}.
\newline\urlprefix\url{http://arxiv.org/abs/1907.05519}

\bibitem{Werthimer1995}
C.~{Werthimer, D.; Ng, D.; Bowyer, S.; Donnelly},
  \href{https://ui.adsabs.harvard.edu/abs/1995ASPC...74..293W/abstract}{{The
  Berkeley SETI Program: SERENDIP III and IV Instrumentation}}, ASPC 74 (1995)
  592.
\newline\urlprefix\url{https://ui.adsabs.harvard.edu/abs/1995ASPC...74..293W/abstract}

\bibitem{Sagan1978}
C.~Sagan,
  \href{https://books.google.co.uk/books/about/Murmurs{\_}of{\_}Earth.html?id=oD90-PBNyr8C{\&}redir{\_}esc=y}{{Murmurs
  of Earth : the Voyager interstellar record}}, Random House, 1978.
\newline\urlprefix\url{https://books.google.co.uk/books/about/Murmurs{\_}of{\_}Earth.html?id=oD90-PBNyr8C{\&}redir{\_}esc=y}

\bibitem{Goldsmith2002}
D.~Goldsmith, T.~C. Owen, {The search for life in the universe}, University
  Science Books, 2002.

\bibitem{Keenan1999}
F.~P. {Irish Astronomical Society.}, M.~E. {Armagh Observatory.}, S.~J.
  {Dunsink Observatory.}, D.~D. Burgess, F.~P. Keenan, M.~E. Phillips, S.~J.
  Rose, D.~D. Burgess,
  \href{https://ui.adsabs.harvard.edu/abs/1999IrAJ...26...87K/abstract}{{The
  Irish astronomical journal.}}, Vol.~26, Irish Astronomical Society, 1999.
\newline\urlprefix\url{https://ui.adsabs.harvard.edu/abs/1999IrAJ...26...87K/abstract}

\bibitem{Vakoch2015}
D.~A. Vakoch, M.~F. Dowd,
  \href{https://www.cambridge.org/gb/academic/subjects/physics/computational-science-and-modelling/drake-equation-estimating-prevalence-extraterrestrial-life-through-ages?format=HB}{{The
  Drake equation : estimating the prevalence of extraterrestrial life through
  the ages}}.
\newline\urlprefix\url{https://www.cambridge.org/gb/academic/subjects/physics/computational-science-and-modelling/drake-equation-estimating-prevalence-extraterrestrial-life-through-ages?format=HB}

\bibitem{Kipping2020}
D.~Kipping, \href{https://www.pnas.org/content/117/22/11995}{{An objective
  Bayesian analysis of life's early start and our late arrival}}, Proceedings
  of the National Academy of Sciences 117~(22) (2020) 11995--12003.
\newblock \href {http://dx.doi.org/10.1073/PNAS.1921655117}
  {\path{doi:10.1073/PNAS.1921655117}}.
\newline\urlprefix\url{https://www.pnas.org/content/117/22/11995}

\bibitem{Busch2011}
M.~W. Busch, R.~M. Reddick, \href{http://arxiv.org/abs/0911.3976}{{Testing SETI
  Message Designs}}\href {http://arxiv.org/abs/0911.3976}
  {\path{arXiv:0911.3976}}.
\newline\urlprefix\url{http://arxiv.org/abs/0911.3976}

\bibitem{Heller2019}
R.~Heller, L.~Kiss, \href{http://arxiv.org/abs/1911.12114}{{Exoplanet Vision
  2050}}\href {http://arxiv.org/abs/1911.12114} {\path{arXiv:1911.12114}}.
\newline\urlprefix\url{http://arxiv.org/abs/1911.12114}

\bibitem{Brin2006}
\href{https://lifeboat.com/ex/shouting.at.the.cosmos}{{Shouting at the
  Cosmos}}.
\newline\urlprefix\url{https://lifeboat.com/ex/shouting.at.the.cosmos}

\bibitem{Zaitsev2017}
\href{http://www.setileague.org/editor/meti.htm}{{Guest Editorial -- Making a
  Case for METI}}.
\newline\urlprefix\url{http://www.setileague.org/editor/meti.htm}

\bibitem{Ball1973}
J.~A. Ball,
  \href{https://linkinghub.elsevier.com/retrieve/pii/0019103573901115}{{The zoo
  hypothesis}}, Icarus 19~(3) (1973) 347--349.
\newblock \href {http://dx.doi.org/10.1016/0019-1035(73)90111-5}
  {\path{doi:10.1016/0019-1035(73)90111-5}}.
\newline\urlprefix\url{https://linkinghub.elsevier.com/retrieve/pii/0019103573901115}

\bibitem{Gertz2016a}
J.~Gertz, \href{http://arxiv.org/abs/1605.05663}{{Reviewing METI: A Critical
  Analysis of the Arguments}}\href {http://arxiv.org/abs/1605.05663}
  {\path{arXiv:1605.05663}}.
\newline\urlprefix\url{http://arxiv.org/abs/1605.05663}

\bibitem{Brin2014}
D.~Brin, \href{https://www.jbis.org.uk/paper/2014.67.8}{{The Search for
  Extraterrestrial Intelligence (SETI) and Whether to send 'Messages' (METI): A
  Case for Conversation, Patience and Due Diligence}}, Journal of the British
  Interplanetary Society 67 (2014) 8--16.
\newline\urlprefix\url{https://www.jbis.org.uk/paper/2014.67.8}

\bibitem{Dumas2014}
S.~Dumas, \href{https://www.jbis.org.uk/paper/2014.67.33}{{The Fear of
  Contact}}, Journal of the British Interplanetary Society 67 (2014) 33--37.
\newline\urlprefix\url{https://www.jbis.org.uk/paper/2014.67.33}

\bibitem{Billingham2014}
J.~{Billingham, J.; Benford},
  \href{http://www.jbis.org.uk/paper.php?p=2014.67.17}{{Costs and Difficulties
  of Interstellar 'Messaging' and the Need for International Debate on
  Potential Risks}}, Journal of the British Interplanetary Society 67 (2014)
  17--23.
\newline\urlprefix\url{http://www.jbis.org.uk/paper.php?p=2014.67.17}

\bibitem{Denning2010}
K.~Denning,
  \href{https://www.sciencedirect.com/science/article/abs/pii/S0094576510000779?via{\%}3Dihub
  https://linkinghub.elsevier.com/retrieve/pii/S0094576510000779}{{Unpacking
  the great transmission debate}}, Acta Astronautica 67~(11-12) (2010)
  1399--1405.
\newblock \href {http://dx.doi.org/10.1016/j.actaastro.2010.02.024}
  {\path{doi:10.1016/j.actaastro.2010.02.024}}.
\newline\urlprefix\url{https://www.sciencedirect.com/science/article/abs/pii/S0094576510000779?via{\%}3Dihub
  https://linkinghub.elsevier.com/retrieve/pii/S0094576510000779}

\bibitem{Denning2011}
K.~Denning,
  \href{https://royalsocietypublishing.org/doi/10.1098/rsta.2010.0230}{{Is life
  what we make of it?}}, Philosophical Transactions of the Royal Society A:
  Mathematical, Physical and Engineering Sciences 369~(1936) (2011) 669--678.
\newblock \href {http://dx.doi.org/10.1098/rsta.2010.0230}
  {\path{doi:10.1098/rsta.2010.0230}}.
\newline\urlprefix\url{https://royalsocietypublishing.org/doi/10.1098/rsta.2010.0230}

\bibitem{Denning2013}
K.~Denning,
  \href{http://link.springer.com/10.1007/978-3-642-35983-5{\_}16}{{Impossible
  Predictions of the Unprecedented: Analogy, History, and the Work of
  Prognostication}}, in: Astrobiology, History, and Society, Advances in
  Astrobiology and Biogeophysics, 2013, pp. 301--312.
\newblock \href {http://dx.doi.org/10.1007/978-3-642-35983-5_16}
  {\path{doi:10.1007/978-3-642-35983-5_16}}.
\newline\urlprefix\url{http://link.springer.com/10.1007/978-3-642-35983-5{\_}16}

\bibitem{Gertz2016b}
J.~Gertz, \href{https://www.jbis.org.uk/paper/2016.69.263}{{Post-Detection SETI
  Protocols {\&} METI: The Time Has Come To Regulate Them Both}}, Journal of
  the British Interplanetary Society 69 (2016) 263--270.
\newline\urlprefix\url{https://www.jbis.org.uk/paper/2016.69.263}

\bibitem{OuterSpaceTreaty}
\href{http://disarmament.un.org/treaties/t/outer{\_}space}{{Disarmament
  Treaties Database: Outer Space Treaty}}.
\newline\urlprefix\url{http://disarmament.un.org/treaties/t/outer{\_}space}

\bibitem{Burgess2016}
C.~Burgess, B.~Vis,
  \href{http://link.springer.com/10.1007/978-3-319-24163-0}{{Interkosmos}},
  Springer International Publishing, Cham, 2016.
\newblock \href {http://dx.doi.org/10.1007/978-3-319-24163-0}
  {\path{doi:10.1007/978-3-319-24163-0}}.
\newline\urlprefix\url{http://link.springer.com/10.1007/978-3-319-24163-0}

\bibitem{Sage2009}
D.~Sage,
  \href{http://journals.sagepub.com/doi/10.1111/j.1467-954X.2009.01822.x}{{Giant
  Leaps and Forgotten Steps: NASA and the Performance of Gender}}, The
  Sociological Review 57~(1{\_}suppl) (2009) 146--163.
\newblock \href {http://dx.doi.org/10.1111/j.1467-954X.2009.01822.x}
  {\path{doi:10.1111/j.1467-954X.2009.01822.x}}.
\newline\urlprefix\url{http://journals.sagepub.com/doi/10.1111/j.1467-954X.2009.01822.x}

\bibitem{Casper1995}
M.~J. Casper, L.~J. Moore,
  \href{http://journals.sagepub.com/doi/10.2307/1389295}{{Inscribing Bodies,
  Inscribing the Future: Gender, Sex, and Reproduction in Outer Space}},
  Sociological Perspectives 38~(2) (1995) 311--333.
\newblock \href {http://dx.doi.org/10.2307/1389295}
  {\path{doi:10.2307/1389295}}.
\newline\urlprefix\url{http://journals.sagepub.com/doi/10.2307/1389295}

\bibitem{Sagan1972}
C.~Sagan, L.~S. Sagan, F.~Drake,
  \href{https://science.sciencemag.org/content/175/4024/881}{{A Message from
  Earth}}, Science 175~(4024) (1972) 881--884.
\newblock \href {http://dx.doi.org/10.1126/SCIENCE.175.4024.881}
  {\path{doi:10.1126/SCIENCE.175.4024.881}}.
\newline\urlprefix\url{https://science.sciencemag.org/content/175/4024/881}

\bibitem{CornellSETI}
\href{https://news.cornell.edu/stories/1999/11/25th-anniversary-first-attempt-phone-et-0}{{It's
  the 25th anniversary of Earth's first attempt to phone E.T. | Cornell
  Chronicle}}.
\newline\urlprefix\url{https://news.cornell.edu/stories/1999/11/25th-anniversary-first-attempt-phone-et-0}

\bibitem{Vakoch1998}
D.~A. Vakoch,
  \href{https://www.sciencedirect.com/science/article/pii/S0094576598000307
  https://linkinghub.elsevier.com/retrieve/pii/S0094576598000307}{{The dialogic
  model: representing human diversity in messages to extraterrestrials}}, Acta
  Astronautica 42~(10-12) (1998) 705--710.
\newblock \href {http://dx.doi.org/10.1016/S0094-5765(98)00030-7}
  {\path{doi:10.1016/S0094-5765(98)00030-7}}.
\newline\urlprefix\url{https://www.sciencedirect.com/science/article/pii/S0094576598000307
  https://linkinghub.elsevier.com/retrieve/pii/S0094576598000307}

\bibitem{Wolverton2004}
M.~Wolverton,
  \href{https://books.google.co.uk/books?id=3eekqPQMlycC{\&}redir{\_}esc=y}{{The
  depths of space : the story of the Pioneer planetary probes}}, Joseph Henry
  Press, 2004.
\newline\urlprefix\url{https://books.google.co.uk/books?id=3eekqPQMlycC{\&}redir{\_}esc=y}

\bibitem{Shuch2011}
H.~P. Shuch,
  \href{http://link.springer.com/10.1007/978-3-642-13196-7}{{Searching for
  Extraterrestrial Intelligence}}, The Frontiers Collection, Springer Berlin
  Heidelberg, Berlin, Heidelberg, 2011.
\newblock \href {http://dx.doi.org/10.1007/978-3-642-13196-7}
  {\path{doi:10.1007/978-3-642-13196-7}}.
\newline\urlprefix\url{http://link.springer.com/10.1007/978-3-642-13196-7}

\bibitem{Quast2018}
P.~E. Quast,
  \href{https://www.cambridge.org/core/product/identifier/S1473550418000290/type/journal{\_}article}{{A
  profile of humanity: the cultural signature of Earth's inhabitants beyond the
  atmosphere}}, International Journal of Astrobiology (2018) 1--21\href
  {http://dx.doi.org/10.1017/S1473550418000290}
  {\path{doi:10.1017/S1473550418000290}}.
\newline\urlprefix\url{https://www.cambridge.org/core/product/identifier/S1473550418000290/type/journal{\_}article}

\bibitem{Michaud1995}
M.~Michaud, {SETI and Diplomacy}, Astronomical Society of the Pacific
  Conference Series 74.

\bibitem{Pitkin1967}
H.~F. Pitkin,
  \href{https://books.google.co.uk/books/about/The{\_}Concept{\_}of{\_}Representation.html?id=AgUVWLswTNEC}{{The
  concept of representation}}, 1967.
\newline\urlprefix\url{https://books.google.co.uk/books/about/The{\_}Concept{\_}of{\_}Representation.html?id=AgUVWLswTNEC}

\bibitem{Goldsmith1990}
D.~Goldsmith,
  \href{https://www.sciencedirect.com/science/article/pii/0094576590901439
  https://linkinghub.elsevier.com/retrieve/pii/0094576590901439}{{Who will
  speak for earth? Possible structures for shaping a response to a signal
  detected from an extraterrestrial civilization}}, Acta Astronautica 21~(2)
  (1990) 149--151.
\newblock \href {http://dx.doi.org/10.1016/0094-5765(90)90143-9}
  {\path{doi:10.1016/0094-5765(90)90143-9}}.
\newline\urlprefix\url{https://www.sciencedirect.com/science/article/pii/0094576590901439
  https://linkinghub.elsevier.com/retrieve/pii/0094576590901439}

\bibitem{Weingart1999}
P.~Weingart,
  \href{https://academic.oup.com/spp/article-lookup/doi/10.3152/147154399781782437}{{Scientific
  expertise and political accountability: paradoxes of science in politics}},
  Science and Public Policy 26~(3) (1999) 151--161.
\newblock \href {http://dx.doi.org/10.3152/147154399781782437}
  {\path{doi:10.3152/147154399781782437}}.
\newline\urlprefix\url{https://academic.oup.com/spp/article-lookup/doi/10.3152/147154399781782437}

\bibitem{Bilder2020}
R.~B. Bilder,
  \href{https://www.cambridge.org/core/product/identifier/S0002930019000861/type/journal{\_}article}{{On
  the Search for Extraterrestrial Intelligence (SETI)}}, American Journal of
  International Law 114~(1) (2020) 87--95.
\newblock \href {http://dx.doi.org/10.1017/ajil.2019.86}
  {\path{doi:10.1017/ajil.2019.86}}.
\newline\urlprefix\url{https://www.cambridge.org/core/product/identifier/S0002930019000861/type/journal{\_}article}

\bibitem{Denning2019}
K.~Denning, S.~J. Dick,
  \href{https://ui.adsabs.harvard.edu/abs/2019BAAS...51g.183D/abstract}{{Preparing
  for the Discovery of Life Beyond Earth}}, Astro2020: Decadal Survey on
  Astronomy and Astrophysics, APC white papers; Bulletin of the American
  Astronomical Society 51~(7) (2019) 183.
\newline\urlprefix\url{https://ui.adsabs.harvard.edu/abs/2019BAAS...51g.183D/abstract}

\bibitem{Dominik2011}
M.~Dominik, J.~C. Zarnecki,
  \href{https://royalsocietypublishing.org/doi/10.1098/rsta.2010.0236}{{The
  detection of extra-terrestrial life and the consequences for science and
  society}}, Philosophical Transactions of the Royal Society A: Mathematical,
  Physical and Engineering Sciences 369~(1936) (2011) 499--507.
\newblock \href {http://dx.doi.org/10.1098/rsta.2010.0236}
  {\path{doi:10.1098/rsta.2010.0236}}.
\newline\urlprefix\url{https://royalsocietypublishing.org/doi/10.1098/rsta.2010.0236}

\bibitem{Anderson2002}
D.~P. Anderson, J.~Cobb, E.~Korpela, M.~Lebofsky, D.~Werthimer,
  \href{http://portal.acm.org/citation.cfm?doid=581571.581573}{{SETI@home: an
  experiment in public-resource computing}}, Communications of the ACM 45~(11)
  (2002) 56--61.
\newblock \href {http://dx.doi.org/10.1145/581571.581573}
  {\path{doi:10.1145/581571.581573}}.
\newline\urlprefix\url{http://portal.acm.org/citation.cfm?doid=581571.581573}

\bibitem{Valentini2014}
L.~Valentini,
  \href{https://www.cambridge.org/core/product/identifier/S1537592714002138/type/journal{\_}article}{{No
  Global Demos, No Global Democracy? A Systematization and Critique}},
  Perspectives on Politics 12~(4) (2014) 789--807.
\newblock \href {http://dx.doi.org/10.1017/S1537592714002138}
  {\path{doi:10.1017/S1537592714002138}}.
\newline\urlprefix\url{https://www.cambridge.org/core/product/identifier/S1537592714002138/type/journal{\_}article}

\bibitem{Gulyas2013}
A.~J. Gulyas,
  \href{https://books.google.co.uk/books/about/Extraterrestrials{\_}and{\_}the{\_}American{\_}Zeitg.html?id=bPxRk{\_}-Wv1wC{\&}redir{\_}esc=y}{{Extraterrestrials
  and the American zeitgeist : alien contact tales since the 1950s}},
  McFarland, 2013.
\newline\urlprefix\url{https://books.google.co.uk/books/about/Extraterrestrials{\_}and{\_}the{\_}American{\_}Zeitg.html?id=bPxRk{\_}-Wv1wC{\&}redir{\_}esc=y}

\bibitem{Rose2009}
J.~Rose,
  \href{http://link.springer.com/10.1057/9780230240902{\_}11}{{Institutionalizing
  Participation through Citizens' Assemblies}}, in: Activating the Citizen,
  Palgrave Macmillan UK, London, 2009, pp. 214--232.
\newblock \href {http://dx.doi.org/10.1057/9780230240902_11}
  {\path{doi:10.1057/9780230240902_11}}.
\newline\urlprefix\url{http://link.springer.com/10.1057/9780230240902{\_}11}

\bibitem{Pal2012}
M.~Pal,
  \href{https://www.semanticscholar.org/paper/The-Promise-and-Limits-of-Citizens{\%}27-Assemblies{\%}3A-and-Pal/3b6cd924c12ccaf79fa86f62e893be22d8175cfb}{{The
  Promise and Limits of Citizens' Assemblies: Deliberation, Institutions and
  the Law of Democracy}} (2012).
\newline\urlprefix\url{https://www.semanticscholar.org/paper/The-Promise-and-Limits-of-Citizens{\%}27-Assemblies{\%}3A-and-Pal/3b6cd924c12ccaf79fa86f62e893be22d8175cfb}

\bibitem{Dula1985}
A.~Dula, \href{https://scholar.smu.edu/til/vol19/iss1/9}{{Private Sector
  Activities in Outer Space}}, International Lawyer 19~(1).
\newline\urlprefix\url{https://scholar.smu.edu/til/vol19/iss1/9}

\bibitem{Denis2020}
G.~Denis, D.~Alary, X.~Pasco, N.~Pisot, D.~Texier, S.~Toulza,
  \href{https://www.sciencedirect.com/science/article/pii/S0094576519313451
  https://linkinghub.elsevier.com/retrieve/pii/S0094576519313451}{{From new
  space to big space: How commercial space dream is becoming a reality}}, Acta
  Astronautica 166 (2020) 431--443.
\newblock \href {http://dx.doi.org/10.1016/j.actaastro.2019.08.031}
  {\path{doi:10.1016/j.actaastro.2019.08.031}}.
\newline\urlprefix\url{https://www.sciencedirect.com/science/article/pii/S0094576519313451
  https://linkinghub.elsevier.com/retrieve/pii/S0094576519313451}

\bibitem{Traphagan2020}
J.~W. Traphagan, {Protocols for Encounter with Extraterrestrials: Lessons from
  the Covid-19 Pandemic}, Journal of the British Interplanetary Society 73
  (2020) 234--238.

\end{thebibliography}


\end{document}